# An Intercomparison Between Divergence-Cleaning and Staggered Mesh Formulations for Numerical Magnetohydrodynamics

By

**Dinshaw S. Balsara[1] and Jongsoo Kim[2]**

(dbalsara@nd.edu, jskim@kao.re.kr)

[1]Physics Department, Univ. of Notre Dame,

[2]Korea Astronomy Observatory

**Abstract**

In recent years, several different strategies have emerged for evolving the magnetic field in numerical MHD. Some of these methods can be classified as divergence-cleaning schemes, where one evolves the magnetic field components just like any other variable in a higher order Godunov scheme. The fact that the magnetic field is divergence-free is imposed post-facto via a divergence-cleaning step. Other schemes for evolving the magnetic field rely on a staggered mesh formulation which is inherently divergence-free. The claim has been made that the two approaches are equivalent. In this paper we cross-compare three divergence-cleaning schemes based on scalar and vector divergence-cleaning and a popular divergence-free scheme. All schemes are applied to the same stringent test problem. Several deficiencies in all the divergence-cleaning schemes become clearly apparent with the scalar divergence-cleaning schemes performing worse than the vector divergence-cleaning scheme. The vector divergence-cleaning scheme also shows some deficiencies relative to the staggered mesh divergence-free scheme. The differences can be explained by realizing that all the divergence-cleaning schemes are based on a Poisson solver which introduces a non-locality into the scheme, though other subtler points of difference are also catalogued. By using several diagnostics that are routinely used in the study of turbulence, it is shown that the differences in the schemes produce measurable differences in physical quantities that are of interest in such studies.

# I) Introduction

In recent years there has been substantial progress in numerical magnetohydrodynamics (MHD). This progress has been spurred by the great utility of this system of equations in the modeling of problems in astrophysics and space physics. A key advance in the field has come from the application of higher order Godunov methodology to numerical MHD. These higher order Godunov schemes were first developed for Euler flows by vanLeer (1979) where they ushered in a new era of robust, accurate and reliable simulation techniques. In order to enjoy the same advantages of accuracy, reliability and robustness in numerical MHD, several groups have developed higher order Godunov schemes for MHD. A brief list includes Brio and Wu (1988), Zachary, Malagoli and Colella (1994), Powell (1994), Dai and Woodward (1994), Ryu and Jones (1995), Roe and Balsara (1996) and Balsara (1998a,b), Balsara (2003). Viewed on a dimension-by-dimension basis, the MHD equations do lend themselves to an interpretation as a system of conservation laws. Since much of the higher order Godunov scheme methodology works best for a system of conservation laws it is, therefore, natural that early efforts focused entirely on the formulation of numerical MHD as yet another system of conservation laws that is amenable to treatment by higher order Godunov methods. In keeping with that view, early efforts also treated the magnetic fields as zone-centered variables.

The above plan seems like an elegant one till one takes a multi-dimensional view of the induction equation. It has the form:

$$\frac{\partial \mathbf{B}}{\partial t} + c\, \nabla \times \mathbf{E} = 0 \qquad (1.1)$$

where **B** is the magnetic field, **E** is the electric field and c is the speed of light. In the specific case of ideal MHD the electric field is given by:

$$\mathbf{E} = -\frac{1}{c}\mathbf{v} \times \mathbf{B} \qquad (1.2)$$

where **v** is the fluid velocity. Eqn. (1.1) is fundamentally different from a conservation law. It does not require the components of the magnetic field to be conserved in a volume-averaged sense. Eqn. (1.1) does predict, via application of Stoke's law, that the magnetic field remains divergence-free. Brackbill and Barnes (1980) and Brackbill (1985) have shown that violating the $\nabla \cdot \mathbf{B} = 0$ constraint leads to unphysical plasma transport orthogonal to the magnetic field as well as a loss of momentum and energy conservation. For that reason, almost all the groups that initially presented higher order schemes for MHD showed at least some awareness of the fact that it is important to preserve the divergence-free aspect of the magnetic field. Thus, Zachary, Malagoli and Colella (1994), Balsara (1998b), Ryu et al (1995) and Kim et al (1999) all suggested that a divergence-cleaning step be used in conjunction with their early higher order Godunov-based MHD schemes. (It might be pointed out that the divergence-cleaning step also destroys strict conservation of magnetic field components, which is the very attribute we sought to preserve via application of conservation laws.) In an effort to accommodate the fact that the early higher order Godunov MHD schemes built up divergence in the magnetic field Powell (1994) took an alternative approach where he tried to modify the MHD equations. He did so by including source terms in the MHD equations that were proportional to $\nabla \cdot \mathbf{B}$. Recently, Dedner et al (2002) extended the so-called "divergence wave" scheme of Powell et al (1994). The advantage of their method is that the divergence correction part, which is composed of evaluating a Riemann problem for a 2X2 system and a scalar source term, can be completely decoupled from the usual MHD solver. The basic idea of the latter two schemes is to propagate any non-zero divergence out of the computational domain or to damp it. We did not include the latter two schemes in our tests.

An alternative line of thought for dealing with the $\nabla \cdot \mathbf{B} = 0$ problem stems from the work of Yee (1966) who built up these ideas for the transport of electromagnetic fields. The basic idea consisted of having a "staggered mesh magnetic field transport algorithm" where the magnetic field components are collocated at the face-centers of each zone and the electric field components are collocated at the edge-centers of the zones. Stoke's law is then applied to eqn. (1.1) to yield a discrete time-update strategy for the face-centered magnetic fields. Because it follows from a discrete version of Stoke's law, the resulting discrete time-update strategy clearly shows that if the magnetic field is divergence-free at the beginning of a time step, it will remain so at the end of the time step. Yee's idea was extended to numerical MHD by Brecht et al (1981). Evans and Hawley (1989) then implemented the staggered mesh algorithm developed by Brecht et al (1981) in their code, coining the term "constrained transport". DeVore (1991) applied the staggered mesh algorithm of Brecht et al (1981) to his FCT scheme, albeit with a more accurate acknowledgement of its antecedents. Similar schemes were developed for higher order Godunov scheme-based MHD by Dai and Woodward (1998), Ryu et al (1998), Balsara and Spicer (1999a), Londrillo and DelZanna (2000) and Balsara (2003). There are different ways to construct the electric field but perhaps the most elegant and economical strategy consists of utilizing the dualism between the components of the Godunov flux and the electric fields, as shown by Balsara and Spicer (1999a). Londrillo and Del Zanna (2000) have shown that a staggered mesh formulation of the form used by the above three references is fundamental to divergence-free evolution of the magnetic field. Toth (2000) made a comparative study of such schemes and found the scheme of Balsara and Spicer (1999a) to be one of the most accurate second order schemes that he tested. Yet, Toth (2000) found that using a zone-centered approach with divergence-cleaning yielded results that were comparable to the staggered mesh formulations for the simple test problems that he used. One is, therefore, led to wonder whether there might be stringent enough test problems where the differences between divergence-cleaning and the staggered mesh formulations may become more apparent? After all, an analysis of the different forms of divergence-cleaning algorithms, which we catalogue in the next section, shows that one should expect differences between them. Furthermore, an inter-comparison of divergence-cleaning algorithms and staggered mesh formulations, which we also undertake in the next section, also shows that we should expect differences. The purpose of this paper is to make the differences apparent by applying all the formulations to the same stringent test problem. The test problem does involve radiative cooling along with the interplay of strong shocks. Such problems are shown to be especially pernicious for certain kinds of schemes for numerical MHD because the colliding, radiative shocks tend form persistent, converging flows that result in a rapid local build up of error in the divergence of the magnetic field. Self-gravitating astrophysical flows also produce persistent, converging flows and should, therefore, be susceptible to the same problems described here.

In Section II we catalogue a few different divergence-cleaning schemes as well as provide a brief recapitulation of the staggered mesh scheme of Balsara and Spicer (1999a). In doing so we also provide inter-comparisons between them, cataloguing insights about these schemes that we have not seen catalogued in a comprehensive way in the literature. In Section III we introduce the test problem and inter-compare numerical results. In Section IV we provide some conclusions.

**II) Divergence-Cleaning Schemes and the Staggered Mesh Scheme of Balsara and Spicer (1999a)**

The MHD equations can be written in an explicitly conservative form as:

$$\frac{\partial}{\partial t}\begin{pmatrix}\rho\\ \rho v_x\\ \rho v_y\\ \rho v_z\\ \mathcal{E}\\ B_x\\ B_y\\ B_z\end{pmatrix}+\frac{\partial}{\partial x}\begin{pmatrix}\rho v_x\\ \rho v_x^2+P+\mathbf{B}^2/8\pi-B_x^2/4\pi\\ \rho v_x v_y-B_x B_y/4\pi\\ \rho v_x v_z-B_x B_z/4\pi\\ \left(\mathcal{E}+P+\mathbf{B}^2/8\pi\right)v_x-B_x(\mathbf{v}\cdot\mathbf{B})/4\pi\\ 0\\ \left(v_x B_y-v_y B_x\right)\\ -\left(v_z B_x-v_x B_z\right)\end{pmatrix}$$

$$+\frac{\partial}{\partial y}\begin{pmatrix}\rho v_y\\ \rho v_x v_y-B_x B_y/4\pi\\ \rho v_y^2+P+\mathbf{B}^2/8\pi-B_y^2/4\pi\\ \rho v_y v_z-B_y B_z/4\pi\\ \left(\mathcal{E}+P+\mathbf{B}^2/8\pi\right)v_y-B_y(\mathbf{v}\cdot\mathbf{B})/4\pi\\ -\left(v_x B_y-v_y B_x\right)\\ 0\\ \left(v_y B_z-v_z B_y\right)\end{pmatrix}+\frac{\partial}{\partial z}\begin{pmatrix}\rho v_z\\ \rho v_x v_z-B_x B_z/4\pi\\ \rho v_y v_z-B_y B_z/4\pi\\ \rho v_z^2+P+\mathbf{B}^2/8\pi-B_z^2/4\pi\\ \left(\mathcal{E}+P+\mathbf{B}^2/8\pi\right)v_z-B_z(\mathbf{v}\cdot\mathbf{B})/4\pi\\ \left(v_z B_x-v_x B_z\right)\\ -\left(v_y B_z-v_z B_y\right)\\ 0\end{pmatrix}=0$$

(2.1)

where $\mathcal{E}=\rho v^2+P/(\gamma-1)+\mathbf{B}^2/8\pi$ is the total energy of the plasma. We assume a Cartesian mesh with edges of size $\Delta x$, $\Delta y$ and $\Delta z$. Also, let the time step be denoted by $\Delta t$. We describe the different schemes for enforcing divergence-free evolution of magnetic field in the ensuing sub-sections:

**II.a) Scalar Divergence-Cleaning (SDC1)**

The scalar divergence cleaning that is catalogued in this sub-section exactly cleans the discrete divergence in physical space after one application but is susceptible to even-odd decoupling. It has been catalogued by several authors. We follow the prescription in Kim et al (1999). In that strategy, at the end of every time step we make the assignment:

$$B_{x, i, j, k} = B_{x, i, j, k} - \frac{\phi_{i+1, j, k} - \phi_{i-1, j, k}}{2\Delta x} \quad ; \quad B_{y, i, j, k} = B_{y, i, j, k} - \frac{\phi_{i, j+1, k} - \phi_{i, j-1, k}}{2\Delta y} \quad ;$$
$$B_{z, i, j, k} = B_{z, i, j, k} - \frac{\phi_{i, j, k+1} - \phi_{i, j, k-1}}{2\Delta z}$$
(2.2)

The vector with components given by $(B_{x, i, j, k}, B_{y, i, j, k}, B_{z, i, j, k})$ satisfies the discrete divergence condition:

$$\frac{B_{x, i+1, j, k} - B_{x, i-1, j, k}}{2\Delta x} + \frac{B_{y, i, j+1, k} - B_{y, i, j-1, k}}{2\Delta y} + \frac{B_{z, i, j, k+1} - B_{z, i, j, k-1}}{2\Delta z} = 0 \tag{2.3}$$

if the scalar $\phi$ satisfies the condition:

$$\frac{\phi_{i+2, j, k} - 2\phi_{i, j, k} + \phi_{i-2, j, k}}{4\Delta x^2} + \frac{\phi_{i, j+2, k} - 2\phi_{i, j, k} + \phi_{i, j-2, k}}{4\Delta y^2} + \frac{\phi_{i, j, k+2} - 2\phi_{i, j, k} + \phi_{i, j, k-2}}{4\Delta z^2} =$$
$$\frac{B_{x, i+1, j, k} - B_{x, i-1, j, k}}{2\Delta x} + \frac{B_{y, i, j+1, k} - B_{y, i, j-1, k}}{2\Delta y} + \frac{B_{z, i, j, k+1} - B_{z, i, j, k-1}}{2\Delta z}$$
(2.4)

The scheme has the following advantages:

A.1) It is exact in physical space. The discrete divergence in eqn. (2.3), measured after a single application of the scheme catalogued in this sub-section, will be exactly zero.

A.2) When using a spectral method, it is fast because one only needs to evaluate FFTs for a single scalar field.

The scheme also has the following disadvantages:

D.1) In eqn. (2.4) $\phi_{i, j, k}$'s are coupled with those at every other cell in the x, y and z-directions. So an original computational domain is divided into eight subdomains, and $\phi$'s are computed in those subdomains separately, which is called even-odd decoupling. In other words, the divergence cleaning step applied to a mesh with $N^3$ zones requires us to make eight independent solutions of the Poisson problem on meshes with $(N/2)^3$ zones each. Eqn. (2.4) is susceptible to even-odd decoupling. (For narrative simplicity we will assume for the rest of this paper that we are dealing with a cubical mesh with N zones on each side. The present comments are, however, more generally applicable to any cuboidal mesh.) It is true that the decoupling enables one to reduce the amount of calculations. However the solution with the decoupling is more susceptible to aliasing due to the reduced sampling.

D.2) It can accommodate to a few different kinds of boundary conditions. In that sense it is somewhat flexible. However, for general boundary conditions, one does not know whether the boundary conditions on $\phi$ in equn. (2.4) should be Dirichlet, Neumann or mixed. In that sense, the scheme is not exactly specifiable for all manner of boundary conditions. Using this method, we can only resolve the question of boundary conditions unambiguously for periodic domains.

D.3) For anything other than periodic domains, and especially for problems with complicated, non-periodic boundaries, it is not possible to use FFT's. In such situations, the method might become slow.

D.4) When using FFT's, aliasing errors cannot be avoided.

D.5) When solving the problem on a parallel machine, one cannot escape the need for all-to-all communication.

D.6) With a lot of effort, this method may perhaps take well to adaptive mesh refinement (AMR), as was done in Powell et al (1999). In that situation, it would certainly be hobbled by the elliptic solution step. As the depth of the AMR hierarchy increases, the divergence-cleaning step will progressively become more computationally expensive.

D.7) The emergence of a non-zero divergence in one local region in a simulation will produce local source terms on the right hand side of eqn. (2.4). The divergence-cleaning involves a Poisson solver, i.e. the solution of an elliptic equation. As a result the local divergence will provide corrections, via eqn. (2.2), to all parts of physical space. Ignoring the effects of periodicity, the corrections will be directly proportional to the right hand side of eqn. (2.4) and inversely proportional to the square of the distance from the region with non-zero divergence. The use of periodic boundaries only introduces more image sources in the problem, thereby degrading the quality of the solution even further. As a result, even though the MHD system is hyperbolic and propagates information locally, the inclusion of the divergence-cleaning step is non-local and produces action at a distance. This can become especially pernicious if persistent divergence-producing structures develop in lines, sheets or volumes, see Jackson (1975), where the fall-off with distance from the source of divergence is even slower.

D.8) The divergence-cleaning is applied every time step and as a result, the damaging effects mentioned above can build up over thousands of time steps.

**II.b) Scalar Divergence-Cleaning (SDC2) (Not Exact but without Even-Odd Decoupling)**

This method is a variant of the method in sub-section II.a and has been catalogued in Balsara (1998b). It does not completely clean the discrete divergence in physical space after one application but is not susceptible to the even-odd decoupling described above. This is done by replacing eqn. (2.4) with:

$$\frac{\phi_{i+1,j,k} - 2\phi_{i,j,k} + \phi_{i-1,j,k}}{\Delta x^2} + \frac{\phi_{i,j+1,k} - 2\phi_{i,j,k} + \phi_{i,j-1,k}}{\Delta y^2} + \frac{\phi_{i,j,k+1} - 2\phi_{i,j,k} + \phi_{i,j,k-1}}{\Delta z^2} = \frac{B_{x,i+1,j,k} - B_{x,i-1,j,k}}{2\Delta x} + \frac{B_{y,i,j+1,k} - B_{y,i,j-1,k}}{2\Delta y} + \frac{B_{z,i,j,k+1} - B_{z,i,j,k-1}}{2\Delta z} \quad (2.5)$$

The remaining equations in sub-section II.a are unchanged. The Poisson problem in eqn (2.5) is solved on a single mesh with $N^3$ zones.

The scheme has the same advantages and disadvantages as the scheme in II.a with the following exceptions:

A.1) It is not susceptible to even-odd decoupling.

D.1) It is not exact in physical space. In fact, successive application of this scheme will cause a different non-zero divergence to be evaluated at each point in real space at the end of each application.

**II.c) Vector Divergence-Cleaning (VDC)**

This scheme is not exact in real (physical) space but it is exact in Fourier space. It was catalogued in Balsara (1998b). It consists of transforming the vector field component-wise into spectral space. Thus, given the three components ($B_{x,i,j,k}$, $B_{y,i,j,k}$, $B_{z,i,j,k}$) at each spatial point (i,j,k) on the mesh, we carry out three three-dimensional FFT's to obtain the field ($B_x(\mathbf{k}), B_y(\mathbf{k}), B_z(\mathbf{k})$) in spectral space. Here $\mathbf{k}$ is the vector of wave-numbers in Fourier space. The field ($B_x(\mathbf{k}), B_y(\mathbf{k}), B_z(\mathbf{k})$) is not divergence-free in Fourier space. However, once we know the B-field in Fourier space, it is easy to correct for a non-zero divergence in Fourier space because the divergence operator in real space transforms to a dot product between $\mathbf{k}$ and ($B_x(\mathbf{k}), B_y(\mathbf{k}), B_z(\mathbf{k})$) in Fourier space. This fact enables one to derive (see Balsara 1998b)

$$B_i(\mathbf{k}) = \sum_{j=1}^{3}\left(\delta_{i,j} - \frac{k_i k_j}{\mathbf{k}^2}\right) B_j(\mathbf{k}) \tag{2.6}$$

The second term in the parenthesis is the component parallel to the wave number. The resultant field on the left side of eqn. (2.6) has a zero scalar product with $\mathbf{k}$. We, therefore, refer to the magnetic field on the left of eqn. (2.6) as being divergence-free in Fourier space. Balsara (1998b) considers further variants of eqn. (2.6) which are more relevant to the discrete form of the divergence operator but those variants do not produce results that are any different from eqn. (2.6). Thus eqn. (2.6) will suffice for our purposes in this work. Transforming eqn. (2.6) it back to real space, therefore, completes the divergence-cleaning step.

The scheme has the following advantages:
A.1) It is not susceptible to even-odd decoupling. This is a major advantage over the SDC1 scheme catalogued in sub-section II.a.
A.2) It is exact up to machine accuracy in Fourier space. In fact, for a mesh with $N^3$ zones, eqn. (2.6) provides $N^3$ divergence-free conditions that hold in Fourier space. Thus there is an equivalence between the information content in real and Fourier space.

A.3) Because of the above two points, this scheme has a major advantage over the SDC2 scheme catalogued in sub-section II.b. The SDC2 scheme avoids the problem of even-odd decoupling, but it is neither exactly divergence-free in physical space nor in Fourier space.

The scheme also has the following disadvantages:

D.1) It is not exact in real space. However, unlike the SDC2 scheme in sub-section II.b, successive application of this scheme will not cause the divergence in physical space to change after every application. In that sense, it is unambiguous.

D.2) It is only applicable to uniform meshes that cover rectangular domains. It cannot be extended to non-Cartesian domains.

D.3) One has to resort to three FFT's instead of one. As a result, it is slower than SDC1 and SDC2.

D.4) When solving the problem on a parallel machine, one cannot escape the need for all-to-all communication. The all-to-all communication needed for the present scheme is thrice as large as the amount of communication needed for SDC1 and SDC2.

D.5) This method cannot be used for adaptive mesh refinement (AMR).

D.6) It might seem that the VDC scheme does not involve a Poisson solver. However, that feeling is illusory. The $1/\mathbf{k}^2$ term in eqn. (2.6) implicitly involves a Poisson solver step. As a result, the VDC scheme also introduces a non-local component into the MHD equations.

D.7) As for the SDC1 and SDC2 schemes, when using FFTs, aliasing errors cannot be avoided.

**II.d) Divergence-Free Staggered Mesh Scheme (SM)**

This scheme is exact in real space. It was described in Balsara and Spicer (1999a). On comparing eqn. (2.1) with a formal conservation law of the form:

$$\frac{\partial \mathbf{U}}{\partial t} + \frac{\partial \mathbf{F}}{\partial x} + \frac{\partial \mathbf{G}}{\partial y} + \frac{\partial \mathbf{H}}{\partial z} = 0 \qquad (2.7)$$

where **F**, **G** and **H** are the fluxes in the x, y and z directions we notice that the flux terms obey the following symmetries:

$$F_7 = -G_6, \quad F_8 = -H_6, \quad G_8 = -H_7 \tag{2.8}$$

The Balsara and Spicer (1999a) scheme is based on realizing that there is a dualism between the fluxes that are produced by a higher order Godunov scheme and the electric fields that are needed in eqn. (1.1). The dualism is put to use by capitalizing on the symmetries in eqn. (2.8). In a higher order Godunov scheme that is spatially and temporally second order accurate, the flux variables are available at the center of each zone's face using a straightforward higher order Godunov scheme. The last three components of the **F**, **G** and **H** fluxes can also be reinterpreted as electric fields in our dual approach. The electric fields are needed at the edge centers as shown in Figure 1 and are to be used to update the face-centered magnetic fields. Thus the Godunov fluxes are directly assigned to the edge centers as follows (eqns. (2.9) to (2.11) should not be viewed as matrix equations):

$$E_{x,\,i,j+1/2,k+1/2}^{n+1/2} = \frac{1}{4c} \begin{pmatrix} H_{7,\,i,j,k+1/2}^{n+1/2} + H_{7,\,i,j+1,k+1/2}^{n+1/2} \\ -\,G_{8,\,i,j+1/2,k}^{n+1/2} - G_{8,\,i,j+1/2,k+1}^{n+1/2} \end{pmatrix} \tag{2.9}$$

$$E_{y,\,i+1/2,j,k+1/2}^{n+1/2} = \frac{1}{4c} \begin{pmatrix} F_{8,\,i+1/2,j,k}^{n+1/2} + F_{8,\,i+1/2,j,k+1}^{n+1/2} \\ -\,H_{6,\,i,j,k+1/2}^{n+1/2} - H_{6,\,i+1,j,k+1/2}^{n+1/2} \end{pmatrix} \tag{2.10}$$

$$E_{z,\,i+1/2,j+1/2,k}^{n+1/2} = \frac{1}{4c} \begin{pmatrix} G_{6,\,i,j+1/2,k}^{n+1/2} + G_{6,\,i+1,j+1/2,k}^{n+1/2} \\ -\,F_{7,\,i+1/2,j,k}^{n+1/2} - F_{7,\,i+1/2,j+1,k}^{n+1/2} \end{pmatrix} \tag{2.11}$$

The magnetic fields are updated by applying a discrete version of Stoke's law to eqn. (1.1) which yields:

$$B^{n+1}_{x, i+1/2, j, k} = B^{n}_{x, i+1/2, j, k} - \frac{c \Delta t}{\Delta y \Delta z} \begin{pmatrix} \Delta z \, E^{n+1/2}_{z, i+1/2, j+1/2, k} - \Delta z \, E^{n+1/2}_{z, i+1/2, j-1/2, k} \\ + \Delta y \, E^{n+1/2}_{y, i+1/2, j, k-1/2} - \Delta y \, E^{n+1/2}_{y, i+1/2, j, k+1/2} \end{pmatrix} \quad (2.12)$$

$$B^{n+1}_{y, i, j-1/2, k} = B^{n}_{y, i, j-1/2, k} - \frac{c \Delta t}{\Delta x \Delta z} \begin{pmatrix} \Delta x \, E^{n+1/2}_{x, i, j-1/2, k+1/2} - \Delta x \, E^{n+1/2}_{x, i, j-1/2, k-1/2} \\ + \Delta z \, E^{n+1/2}_{z, i-1/2, j-1/2, k} - \Delta z \, E^{n+1/2}_{z, i+1/2, j-1/2, k} \end{pmatrix} \quad (2.13)$$

$$B^{n+1}_{z, i, j, k+1/2} = B^{n}_{z, i, j, k+1/2} - \frac{c \Delta t}{\Delta x \Delta y} \begin{pmatrix} \Delta x \, E^{n+1/2}_{x, i, j-1/2, k+1/2} - \Delta x \, E^{n+1/2}_{x, i, j+1/2, k+1/2} \\ + \Delta y \, E^{n+1/2}_{y, i+1/2, j, k+1/2} - \Delta y \, E^{n+1/2}_{y, i-1/2, j, k+1/2} \end{pmatrix} \quad (2.14)$$

This scheme has the following advantages:

A.1) It is exactly divergence-free in real space. This exactness holds true in a very strong sense. Thus if the method is exactly divergence-free in a discrete sense in each zone in physical space then it can be shown that the method is divergence-free at every point in physical space. This is true because one can use the divergence-free reconstruction of vector fields that has been presented in Balsara (2001) to reconstruct the divergence-free magnetic field at all points in space. Balsara (2003) amplifies on this point much further, showing that an MHD scheme that relies on divergence-free reconstruction of vector fields has several advantages over one that does not use these concepts.

A.2) One may assert that on a mesh with $N^3$ zones, this scheme corresponds to naturally imposing $N^3$ divergence-free conditions in physical space. The VDC scheme in sub-section II.c can be viewed as imposing $N^3$ divergence-free *discrete* conditions in Fourier space. However, because of the previous point we realize that the present scheme furnishes a divergence-free representation of the magnetic field at all points in the computational domain in a *continuous* sense. Thus the SM scheme can be made to yield a stronger divergence-free constraint, i.e. one that is valid at any point in the computational domain, as shown by Balsara (2003). The VDC scheme only constrains the magnetic field to be divergence-free at discrete points in Fourier space. Thus there is indeed a real difference between the two schemes. As a result, on sufficiently difficult problems, we do expect to see differences.

A.3) It has been shown to extend naturally to AMR in Balsara (2001).

A.4) On parallel machines, it requires minimal inter-processor communication.

A.5) There are no even-odd decoupling issues to deal with.

A.6) The issue of aliasing does not arise because the method does not use FFTs. In that sense, this scheme has an advantage over the VDC scheme in sub-section II.c.

A.7) Unlike the SDC schemes, this method does not involve Poisson solvers and so it is totally local in its influence.

Table I provides a comparative synopsis of the points made in this section.

**III) Test Problem**

**III.a) Description of the Test Problem**

This present test problem derives from attempts to study the evolution of the turbulent interstellar medium. It has been studied in MacLow et al (2003) and subsequent papers in that series, see Balsara et al (2003) and Kim et al (2003). It simulates a patch of the turbulent interstellar medium (ISM) of a galaxy. Because the ISM is rendered turbulent by strong supernova explosions, the problem consists of having strong point-wise explosions in the medium. To a first approximation, these explosions occur at random points in the Galaxy. As a result, we initiate supernova explosions at random points in the computational domain. The physics of the problem is described in the papers cited above. In this paper, we describe the problem in its computational aspects and units. The three dimensional computational domain covers [-0.1, 0.1]X[-0.1, 0.1]X[-0.1, 0.1] and $128^3$ zones are used in the simulation. Periodic boundary conditions are used. The medium is initially static and has $\rho = 1.0$, $P = 0.3$ and $B_x = 0.056117$. The remaining components of the magnetic field and all three components of the velocity are initially zero. The ratio of the specific heats in the gas is 5/3. Our procedure for initializing supernovae consists of first identifying a random location in the computational domain and resetting the pressure to 13649.6 in the zones that are within a radial distance of 0.005 from the chosen point. Supernovae are initialized every 0.00125 units of simulation time. The sequence of x, y, z positions at which supernovae are initialized is listed in Table II for the first 0.035688 units of simulation time, by which time interesting differences arise between the schemes being tested. Interstellar heating and cooling were incorporated and the detailed description of the physical processes is given in Mac Low et al (2003) and references therein. Because of the energetic input in the supernovae explosions, the heating and cooling are needed to help the system maintain temperature equilibrium over long intervals of time. The results presented here were verified to be independent of the details of heating and cooling processes.

The magnetic field that is initialized in this problem has a magnetic pressure that is much smaller than the gas pressure. Successive explosions generate vorticity and helicity as they interact with the turbulence left behind by the prior set of explosions, as shown in Balsara, Benjamin and Cox (2001). This builds up helicity which, in turn, causes an increase in the magnetic energy. Because of the significant changes in field topology and field energy, this is a stringent test problem for the schemes being tested. We will, therefore, focus on the evolution of the magnetic energy as a function of time, images of the magnetic pressure at different representative times as well as on the histograms and spectra of the magnetic field. These are very standard diagnostics that are used in the turbulence community for analyzing the results of turbulence simulations.

The problem was simulated using an algorithm that is second order accurate in space and time. Temporal accuracy is obtained by using a two-stage, i.e. predictor-corrector type scheme. Each stage uses piecewise-linear interpolation on the primitive variables. The van Leer limiter is used for the interpolation of the density and magnetic field variables. Pressure and velocity variables are interpolated using a minmod limiter. Through extensive experimentation Balsara has found that this results in rather nice results in MHD calculations at a very low cost, see Balsara (2003). The interpolation scheme is coupled with an HLL-type Riemann solver to yield a spatially second order accurate scheme. Pressure positivity is ensured using the formulation in Balsara and Spicer (1999b). The algorithm described here is one of several algorithms implemented in Balsara's RIEMANN framework for computational astrophysics. While it is not as accurate as the method described in Balsara (1998b), its twin advantages are robustness and speed without relinquishing second order accuracy. The second order accuracy of the present algorithm is catalogued in Balsara et al (2003).

**III.b) Description of the Results**

**III.b.1) Energy Evolution of the Magnetic Field**

Fig 2 shows the evolution of the magnetic energy as a function of time for the four different schemes that are inter-compared. The divergence-cleaning schemes, SDC1, SDC2 and VDC were applied to the simulation at every time step and the results are shown in Figs. 2.a, 2.b and 2.c respectively. The result from the SM scheme is shown in Fig. 2.d. From Fig 2.a we see that the SDC1 scheme shows a chaotic and very rapid rise in the magnetic energy. Even after the first few supernova explosions, there is a dramatic increase in the magnetic energy. This increase sets in as soon as we have a situation where one explosion interacts with the complicated magnetic field structure left behind by a prior explosion. This point is made graphically in Fig. 3 which shows an illustrative slice plane at a time of 0.005948 for the four schemes being tested. The SDC1 scheme is applied after every time step and does cause the magnetic field to be restored to its divergence-free state after each application of eqn. (2.2). However, SDC1 is also susceptible to even-odd decoupling. As a result, even though the divergence is zero in each zone in the sense of eqn. (2.3), each zone is decoupled from the zones that surround it in the divergence-cleaning step. For that reason, even though the scheme treats isolated explosions just fine (as shown by Balsara (1998b) and Toth (2000)), when explosions interact and generate small-scale structure in the process, there is a dramatic increase in the magnetic energy. At late times, the supernovae inevitably explode in a field of relic turbulence left behind by prior explosions. Each of the spikes in Fig 2.a at late times corresponds to the interaction of a new supernova explosion with the ambient turbulence. From Fig 2.b we see that the SDC2 scheme also shows a chaotic and rather rapid rise in the magnetic energy. From Fig. 2.c we see that the VDC scheme also shows some fluctuations. The SM scheme, for which the temporal evolution of the magnetic energy is shown in Fig 2.d, is the only scheme for which the magnetic energy grows in a monotone fashion without any strong fluctuations. When two remnants collide they produce a converging flow that stagnates at the point of collision. Cooling tends to abet the formation of very persistent, converging flows over very small length scales (as does gravity in self-gravitating flows). Due to the formation of persistent, converging flows we have rapid, localized build up of non-zero divergence which the SDC1, SDC2 and VDC schemes are incapable of removing. Even the SDC1 scheme, which makes the discrete divergence mathematically zero in real space, is not without aliasing errors due to even-odd decoupling. The SM scheme, which doesn't rely on divergence-cleaning, faithfully describes the field evolution in such situations.

It is worth pointing out that the fluctuations in the VDC scheme are small enough that one could perhaps have extracted the rate of growth of the magnetic field by averaging over the fluctuations. Furthermore, once the fluctuations subside, the scheme produces the same amount of growth in magnetic energy as the SM scheme. We have also run all four schemes for a length of time that it is almost seven times longer than the times reported here. The magnetic energy in the VDC and SM schemes reaches the same final level once the fluctuations in the VDC scheme have subsided. This provides an important cross-check for the two schemes. The cross-check is important because spectral schemes, see for example Zeinecke, Politano and Pouquet (1998), invariably use the VDC scheme and have been successfully applied to incompressible dynamo simulations. The magnetic energy in the SDC1 and SDC2 schemes undergoes explosive growth and saturates at a level that is an order of magnitude larger than the VDC and SM schemes. We, therefore, identify the SDC1 and SDC2 schemes as being unsuitable for dynamo applications. It must also be noted that had the problem not been so severe, as is the case for incompressible and moderately compressible dynamo simulations, the VDC scheme would have performed just fine.

**III.b.2) Magnetic Structures**

The evolution of magnetic energy that was discussed in the previous two paragraphs gives us a clue that something may be wrong with the SDC1/2 schemes and the VDC scheme. But it does not make the reason for the fluctuations graphically clear. That is done in Figs. 3 and 4. Fig 3.a images the logarithm of the magnetic pressure in a selected slice plane showing the shells a brief time after they have interacted, i.e. at a simulation time of 0.005948 , for the SCD1 scheme. Figs. 3.b, 3.c and 3.d show the logarithm of the magnetic pressure in the same slice plane at the same time for the SDC2, VDC and SM schemes respectively. (Taking the logarithm of the magnetic pressure only helps in bringing out the contrasts that would otherwise be missed.) From Fig. 3.a we expect that the interacting shells in the three-dimensional simulation intersect in a plane, which is shown as a line in the slice plane. We can also see that the spurious fluctuations are most prominent in the portion of the interior of each interacting shell that is closest to the plane of interaction and decrease with distance from that plane. Since the plane in which the shells intersect is also the region where the non-local divergence-cleaning is maximally applied (see point D.7 in Section II.a), we expect that the zones around that region will be maximally influenced by the divergence-cleaning step in eqn. (2.2). This expectation is exactly borne out in Fig. 3.a. We can also see that the interior of the isolated shell also has small fluctuations in it. However, the fluctuations are not so pronounced that one would focus on them unless one anticipated them in advance. Fig 3.b shows that SDC2 shares many of the same traits as SDC1. We do, however, see that the pronounced banded structure that is visible in the lower interacting shell of Fig 3.a is not present in Fig 3.b. This is symptomatic of the fact that SDC2 is free of even-odd decoupling while SDC1 is not. Thus we see that even though SDC2 does not clean the divergence entirely in one step, it has some better properties because it does not suffer from some of the other limitations of SDC1. Fig. 3.c shows the same slice plane from the VDC scheme. We see that it too has unphysical fluctuations in the magnetic energy around the plane where the two shells interact. We see from Fig. 3.c that spurious fluctuations in the VDC scheme are most prominent in the portion of the interior of each interacting shell that is closest to the plane of interaction and decrease with distance from that plane. Thus VDC, SDC1 and SDC2 all display the same kinds of spurious fluctuations. This clearly bears out the point made in D.6 of Section II.c where we pointed out that the VDC scheme also has a non-local character implicitly built into it because of the use of a Poisson solver. Thus, even though the VDC scheme exactly imposed $N^3$ divergence-free conditions in spectral space (just as the SDC1 scheme imposes $N^3$ divergence-free conditions in real space) that is still not sufficient in ensuring that it obtains a physical result. The reason invariably stems from the non-local aspect of the SDC1, SDC2 and VDC schemes. The above three schemes are also not free of aliasing effects. Fig. 3.d shows the same slice plane for the SM scheme. The SM scheme is free of the unphysical fluctuations in the magnetic energy that emerged in the SDC1, SDC2 and VDC schemes. Moreover, we see that the interior of the isolated supernova shell in Fig 3.d is entirely free of small-scale fluctuations, illustrating the usefulness of a divergence-free scheme that is totally local. We also remind the reader of points A.1 and A.2 in Section II.d which show that a scheme that is divergence-free in a discrete sense has the special property that its magnetic field can be reconstructed in a divergence-free fashion at every point in physical space. This is a property that does not hold true for the VDC scheme.

It is also interesting to ask what happens at a somewhat later time when several supernova shells have begun to intersect? Figs. 4.a, 4.b, 4.c and 4.d show the same slice plane as in Fig 3 at a later time of 0.0011896 for the SDC1, SDC2, VDC and SM schemes respectively. We see that the fluctuations in Fig. 4.a for the SDC1 scheme have grown to the point where they are on par with the compressed field in the individual supernova shells! The fluctuations in Fig. 4.b for the SDC2 scheme are also unacceptably large but not so large as to swamp out the shell boundaries. From Fig. 4.c we see that the magnetic field fluctuations in the VDC scheme are also quite large. They are, however, better localized and, for the most part, do not exceed the strength of the magnetic field in the shells. From Fig. 4.d we see that the SM scheme is the only scheme which produces interacting shells with clean interiors!

**III.b.3) Statistics of the Magnetic Field**

In the previous two paragraphs we showed that the turbulent structures that are produced by the SDC1, SDC2 and VDC schemes can have some faulty elements in them. In turbulence simulations it is also of interest to study statistics and spectra. It is, therefore, worthwhile to ask whether differences show up in the statistical measures that are traditionally used in turbulence studies? In view of the computational focus of this paper, we focus on histograms of the magnitude of the magnetic field, which is but one of many popular diagnostics that are used in turbulence research. Fig 5.a shows the histograms for the SDC1 scheme at times of 0.004758, 0.005948 and 0.008327 . The times have also been shown by the vertical lines in Fig. 2.a and straddle the first strong fluctuation in magnetic energy that sets in when the first two shells interact. The horizontal axis in the histogram shows the strength of the magnetic field and the vertical axis shows the fraction of the computational volume that has that field strength. Fig 5.b shows the histograms for the SDC1 scheme at times of 0.03093, 0.03212 and 0.03331 . The times have also been shown by the vertical lines in Fig. 2.a and straddle a strong fluctuation in magnetic energy that occurs much later. Fig. 5.c shows the same times as Fig. 5.a for the SM scheme. Fig. 5.d shows the same times as Fig. 5.b for the SM scheme. From Fig. 5.a we see that when the magnetic energy spikes upwards in Fig. 2.a at a time of 0.005948 the histogram undergoes a shift to the right indicating that when the shells intersect several zones in the computation acquire strong magnetic fields. From Fig. 5.a we also see that when the magnetic energy has subsided in Fig. 2.a at a time of 0.008327 the histogram has resumed a shape quite similar to its original shape at a time of 0.004758 . From Fig. 5.c we see that the SM scheme does not suffer from this deficiency. Fig 5.b corresponds to the SDC1 scheme at a later time when the magnetic energy in Fig. 2.a undergoes a similar, albeit smaller, spike. Again we observe that the histogram of the magnetic field for the SDC1 scheme undergoes a substantial shift. The histogram for the SM scheme in Fig. 5.d does not undergo a similar shift. The shift of the histograms is clearly a consequence of the FFT-based elliptic solver which causes local information to propagate globally. We also see from Fig. 5 that the histograms broaden with increasing time. This is an expected result for the present class of turbulence simulations. Since histograms are one of the most popular ways of gathering statistics from a turbulent simulation, we have convincingly demonstrated in this paragraph that one is apt to extract spurious results from an inadequate scheme for numerical MHD.

### III.b.4) Spectra of the Magnetic Field

Spectral analysis is another very powerful and well-used strategy for extracting diagnostics from a turbulence simulation. In Fig. 6 we show power spectra for the magnetic field at a time of 0.035688 units of simulation time for all four schemes. By this point, all points in the simulation have been processed at least once by the supernova explosions. We see that the SDC1 and SDC2 schemes produce a considerable amount of small-scale power. This is consistent with the spurious structures they produced in Figs. 3 and 4 and the peculiar shifts in the histograms in Fig. 5. From Fig. 6 we see that SDC1 produces the worst spectrum because of its susceptibility to even-odd decoupling. The spectrum from SDC2 is marginally better though the SDC2 scheme also suffers from the deficiency that it is not exact in real space. The spectrum from the VDC scheme is quite good though it too shows some spurious energy on the smallest scales. The SM scheme alone produces spectra with well-tempered dissipation characteristics on the smallest scales.

### IV) Conclusions

Based on the work presented here we offer the following conclusions:

1) We have cross-tested the SDC1, SDC2 and VDC (divergence-cleaning) schemes and and the SM (divergence-free) scheme for numerical MHD. We have found that there are important differences in the results if one looks closely enough with a rich-enough set of diagnostic tools. The differences become quite pronounced when the physical problem becomes rather stringent. This has been demonstrated by applying the four schemes to the same test problem of supernova-induced MHD turbulence in the ISM.

2) The scalar divergence cleaning strategies, like those examined by Toth (2000), show spikes in the magnetic energy on stringent enough problems. The physical reason for these spikes is explained. The vector divergence cleaning method also shows some spurious energetic fluctuations but it does, at least, converge to the right level of energy after the fluctuations have died out.

3) The problems with the SDC1, SDC2 and VDC schemes are properly explained as being a consequence of the non-locality introduced into numerical MHD by the divergence-cleaning strategy. The schemes also suffer from aliasing errors. The SDC1 scheme also suffers from even-odd decoupling.

4) Examination of the magnetic field structures shows that the difference between the divergence-cleaning and divergence-free schemes will become especially pronounced when persistent linear, sheet-like and volumetric structures form in the flow that have a tendency to generate divergence. The physical reasoning for that is given in the text.

5) Statistical and spectral analysis also show the deficiency of the divergence-cleaning schemes, thereby limiting their utility in any numerical study of MHD turbulence.

6) Spectral analysis of the SM scheme on some of the most stringent problems that we have been able to design shows that the method is not susceptible to unbounded growth of spurious oscillations.

7) We have shown that SDC1 and SDC2 produce spurious energetics, structures, statistics and spectra – the four mainstays of turbulence studies. From this we conclude that the SDC1 and SDC2 schemes are really not suitable for turbulence studies. The VDC scheme shows some deficiencies. The SM scheme is the only scheme tested that does not show deficiencies on any of the fronts on which we tested the schemes, showing that it is uniquely well-suited for turbulence studies.

**Acknowledgements** : Balsara and Kim acknowledge support via NSF grants R36643-7390002 and 005569-001. Kim was also supported by KOSEF through Astrophysical Research Center for the Structure and Evolution of the Cosmos (ARCSEC).

**Figures**

Figure 1 shows the collocation of the magnetic fields at the control volume's faces and the collocation of electric fields at the control volume's edges for the SM scheme.

Fig 2 shows the evolution of the magnetic energy as a function of time for the four different schemes that are inter-compared. The divergence-cleaning schemes, SDC1, SDC2 and VDC were applied to the simulation at every time step and the results are shown in Figs. 2.a, 2.b and 2.c respectively. The result from the SM scheme is shown in Fig. 2.d.

Fig. 3 shows an illustrative slice plane at a time of 0.005948 for the four schemes being tested. The images show the logarithm of the magnetic pressure in a selected slice plane showing the shells a brief time after they have interacted. Figs. 3.a (upper left), 3.b (upper right), 3.c (lower left) and 3.d (lower right) correspond to the SDC1, SDC2, VDC and SM schemes respectively.

Figs. 4.a (upper left), 4.b (upper right), 4.c (lower left) and 4.d (lower right) show the same slice plane as in Fig 3 at a later time of 0.0011896 for the SDC1, SDC2, VDC and SM schemes respectively.

Fig 5.a shows the histograms of the magnitude of the magnetic field for the SDC1 scheme at times of 0.004758, 0.005948 and 0.008327 . Fig 5.b shows the histograms for the SDC1 scheme at times of 0.03093, 0.03212 and 0.03331 . Figs 5.c and 5.d show the same data as Figs. 5.a and 5.b respectively for the SM scheme.

Fig. 6 shows the power spectra for the magnetic field at a time of 0.035688 units of simulation time for all four schemes.

**Table I**

| Attribute/Scheme | SDC1 | SDC2 | VDC | SM |
|---|---|---|---|---|
| Exactly divergence-free? | Yes (in physical space) | No | Yes (in spectral space) | Yes (in physical space) |
| Is the method fast? | Yes | Yes | No | Yes |
| Even-odd decoupling? | Yes | No | No | No |
| Non-periodic Boundaries? | Maybe | Maybe | No | Yes |
| Aliasing Errors? | Yes | Yes | Yes | No |
| All-to-all Communication? | Yes | Yes | Yes | No |
| Uses (non-local) elliptic part? | Yes | Yes | Yes | No |
| Usable in AMR? | Maybe | Maybe | No | Yes |
| Accumulative build-up of error? | Yes | Yes | Yes | No |

**Table II**

| # | X | Y | Z | # | X | Y | Z |
|---|---|---|---|---|---|---|---|
| 1 | 7.825E-07 | 1.315E-02 | 7.556E-02 | 2 | -5.413E-02 | -4.672E-02 | -7.810E-02 |
| 3 | -3.211E-02 | 6.793E-02 | 9.346E-02 | 4 | -6.165E-02 | 5.194E-02 | -1.690E-02 |
| 5 | 5.346E-03 | 5.297E-02 | 6.711E-02 | 6 | 7.698E-04 | -6.165E-02 | -9.331E-02 |
| 7 | 4.174E-02 | 6.867E-02 | 5.889E-02 | 8 | 9.304E-02 | -1.538E-02 | 5.269E-02 |
| 9 | 9.196E-03 | -3.460E-02 | -5.840E-02 | 10 | 7.011E-02 | 9.103E-02 | -2.378E-02 |
| 11 | -7.375E-02 | 4.746E-03 | -2.639E-02 | 12 | 3.653E-02 | 2.470E-02 | -1.745E-03 |

| 13 | 7.268E-03 | -3.683E-02 | 8.847E-02 | 14 | -7.272E-02 | 4.364E-02 | 7.664E-02 |
| 15 | 4.777E-02 | -7.622E-02 | -7.250E-02 | 16 | -1.023E-02 | -9.079E-03 | 6.056E-03 |
| 17 | -9.534E-03 | -4.954E-02 | 5.162E-02 | 18 | -9.092E-02 | -5.223E-03 | 7.374E-03 |
| 19 | 9.138E-02 | 5.297E-02 | -5.355E-02 | 20 | 9.409E-02 | -9.499E-02 | 7.615E-02 |
| 21 | 7.702E-02 | 8.278E-02 | -8.746E-02 | 22 | -7.306E-02 | -5.846E-02 | 5.373E-02 |
| 23 | 4.679E-02 | 2.872E-02 | -8.216E-02 | 24 | 7.482E-02 | 5.545E-02 | 8.907E-02 |
| 25 | 6.248E-02 | -1.579E-02 | -8.402E-02 | 26 | -9.090E-02 | 2.745E-02 | -5.857E-02 |
| 27 | -1.130E-02 | 6.520E-02 | -8.496E-02 | 28 | -3.186E-02 | 3.858E-02 | 3.877E-02 |
| 29 | 4.997E-02 | -8.524E-02 | 5.871E-02 | 30 | 8.455E-02 | -4.098E-02 | -4.438E-02 |

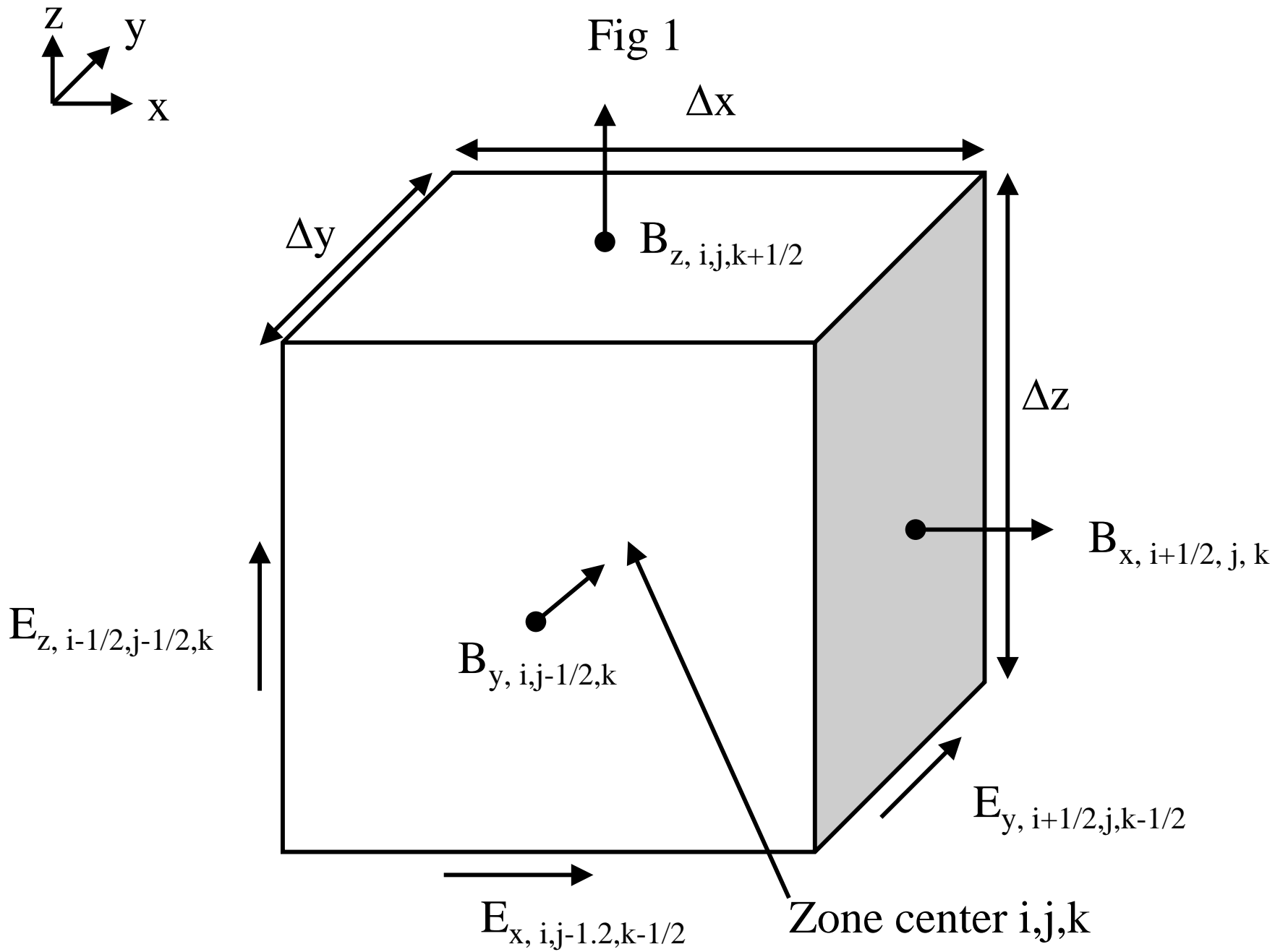

Fig 1

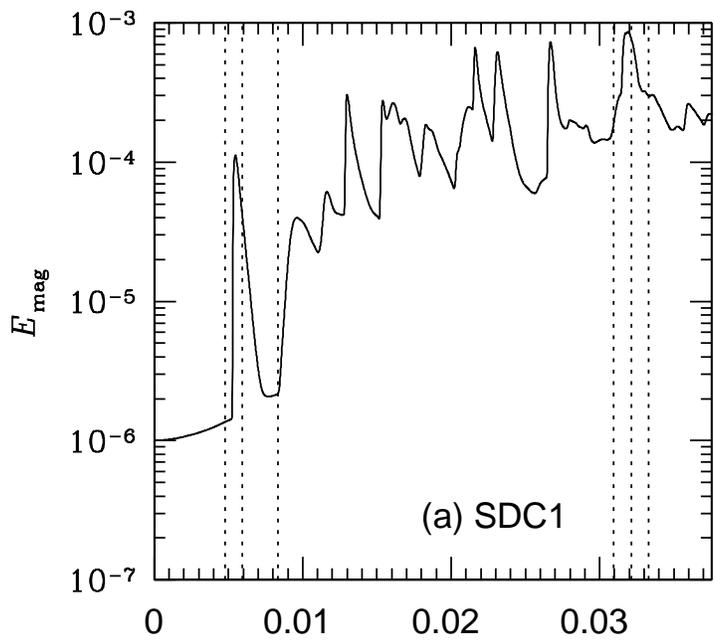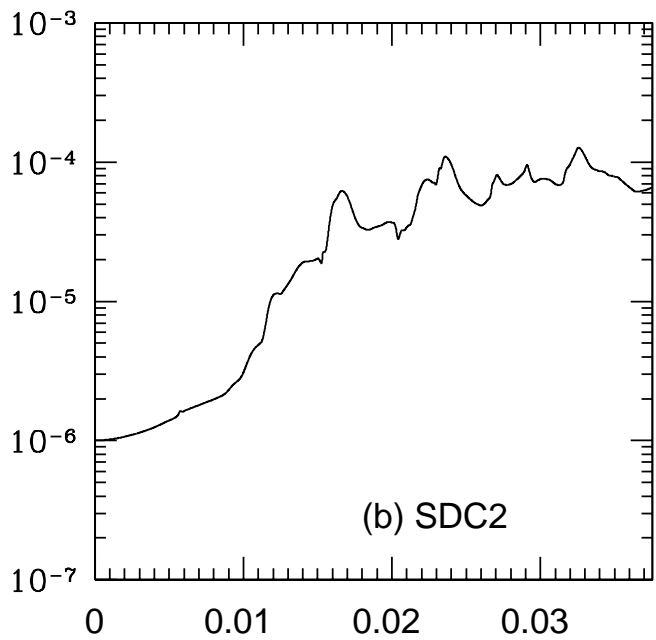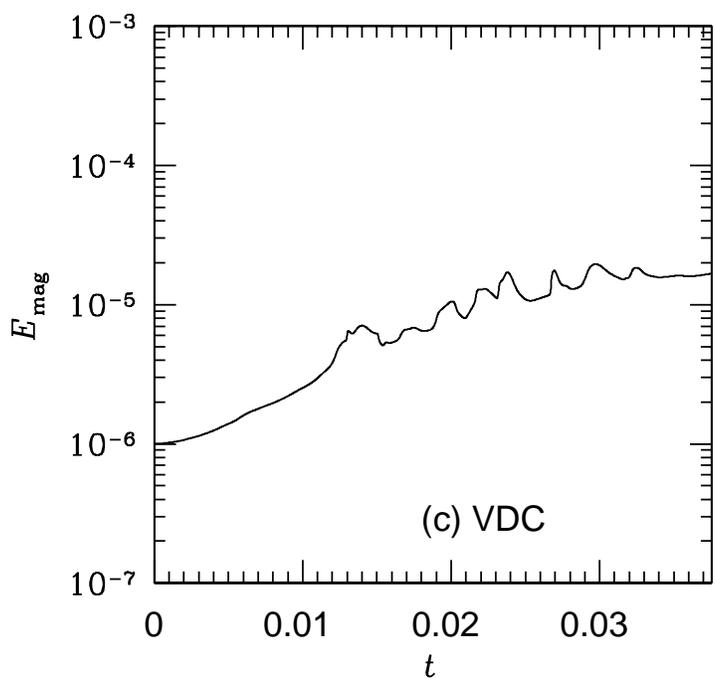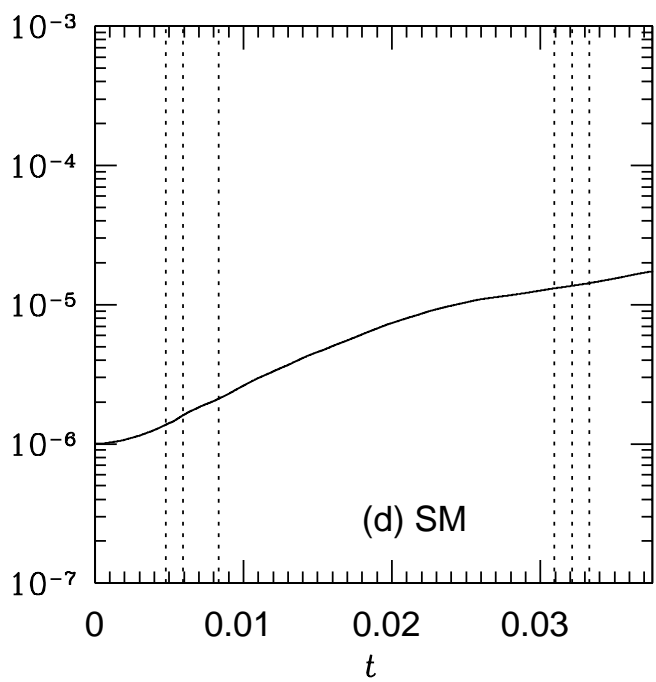

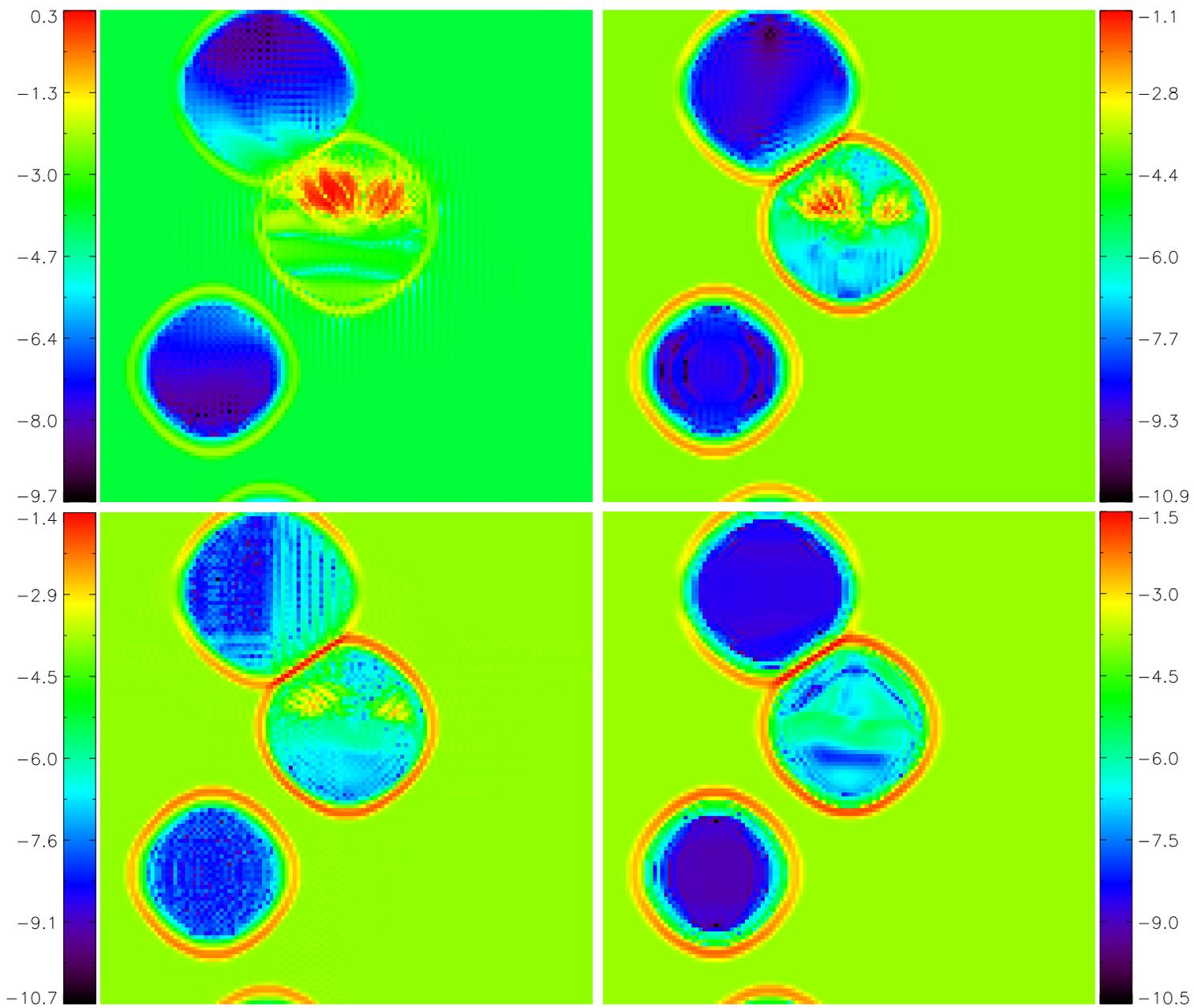

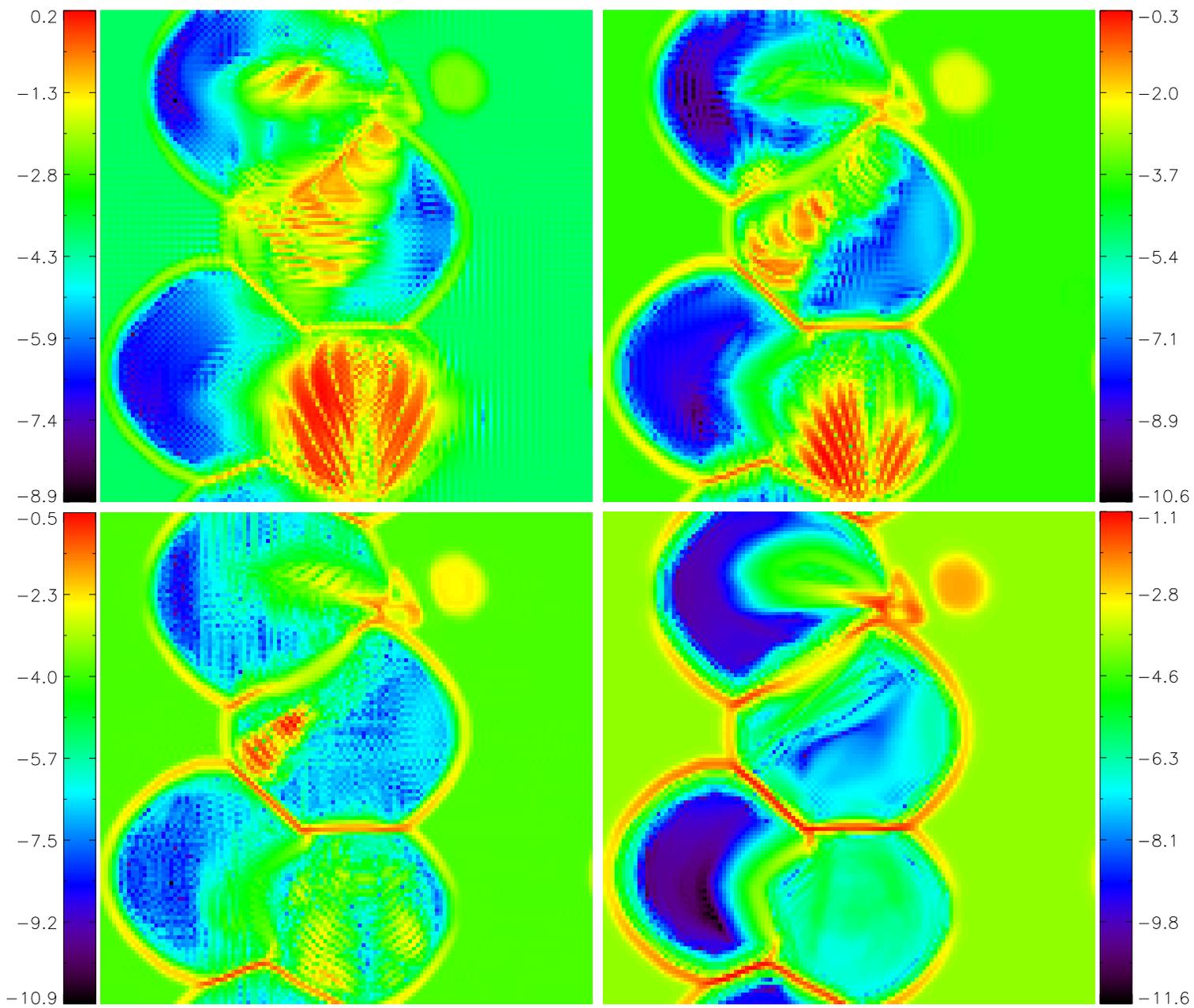

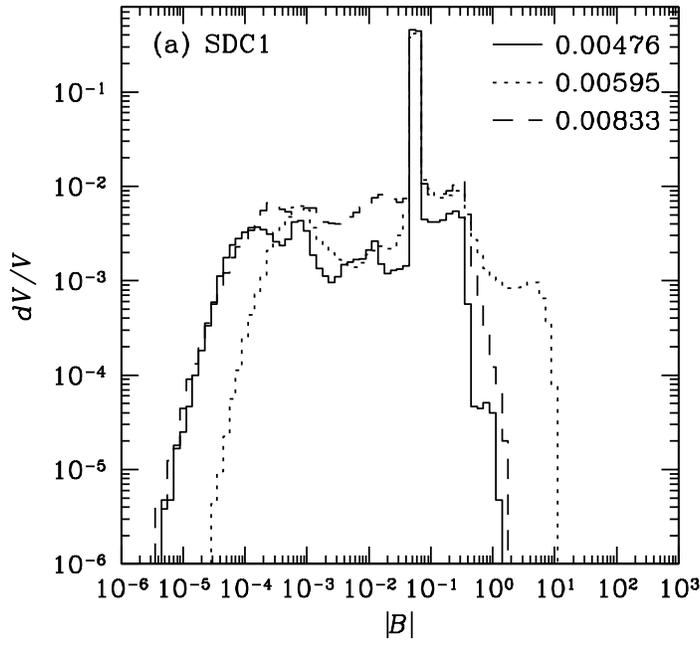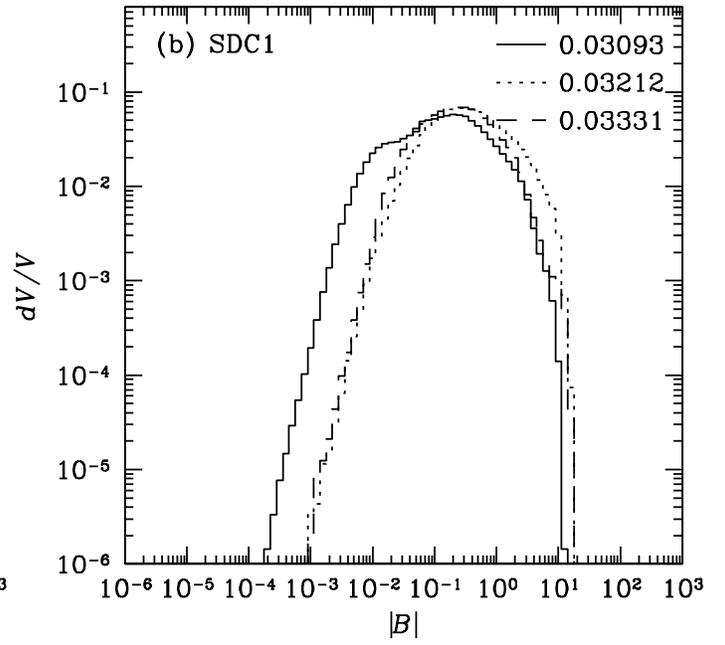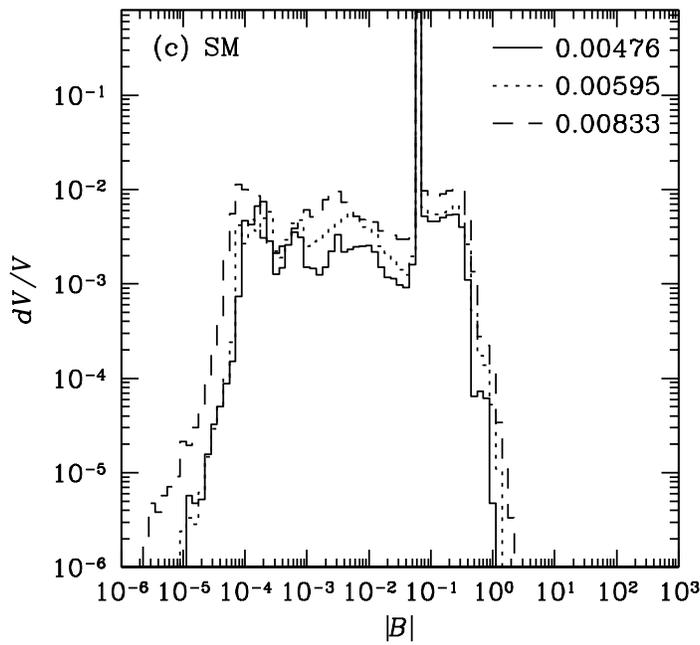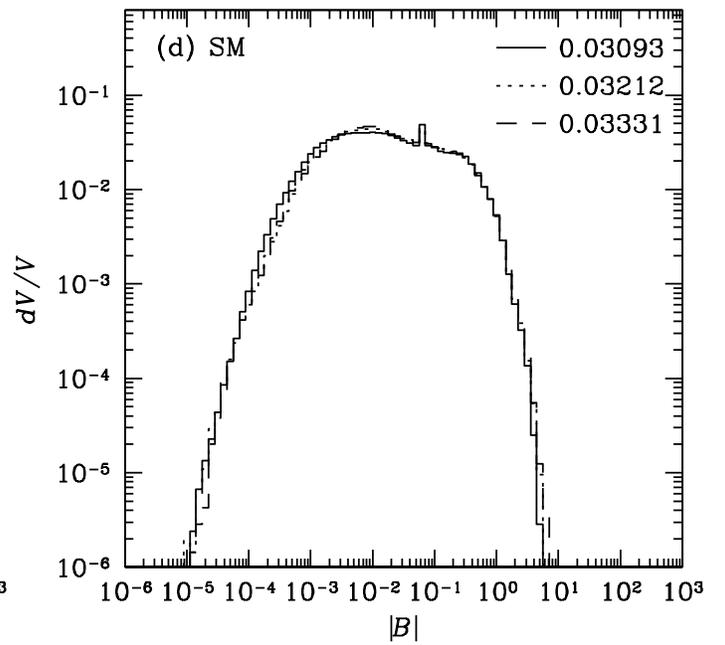

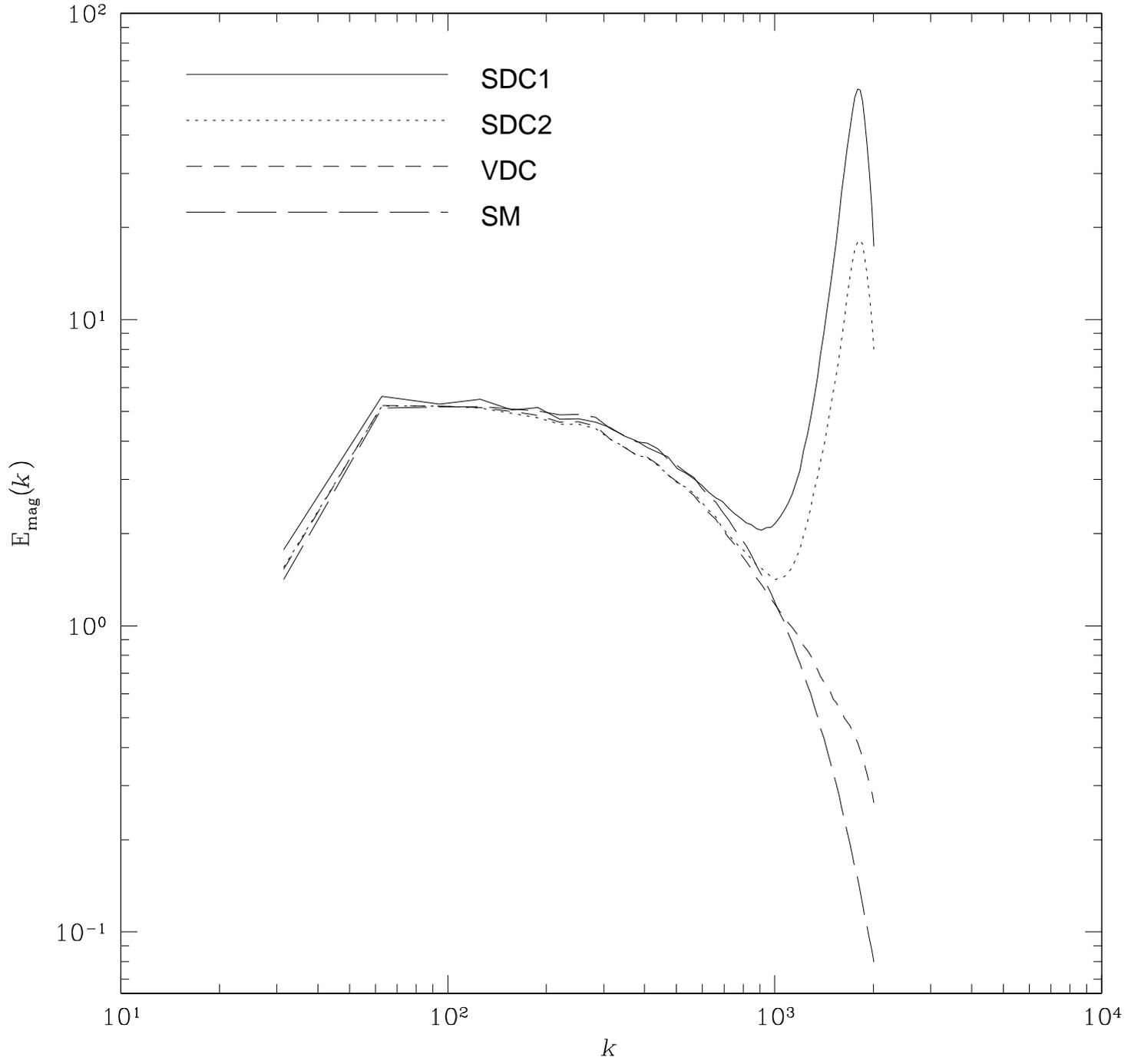